\documentclass[aps,pra%,floatfix,12
]{revtex4-2}
\usepackage{amsfonts}
\usepackage{amsmath}
\usepackage{amsthm}
\usepackage{amssymb}
\usepackage{graphicx}
\usepackage{appendix}
\usepackage{hyperref}
\usepackage{textcomp}
\usepackage{multirow}
\usepackage{color}
\usepackage{bm}
\usepackage{braket}
\usepackage{subfigure}
\usepackage{setspace}

\begin{document}

\title{Interactions between fractional solitons in bimodal fiber cavities}
\author{Thawatchai Mayteevarunyoo$^{1}$, and Boris A. Malomed$^{2,3}$}
\address{$^{1}$Department of Electrical and Computer Engineering, Faculty of
	Engineering, Naresuan University, Phitsanulok 65000, Thailand\\
	$^{2}$Department of Physical Electronics, School of Electrical Engineering,
	Faculty of Engineering, and the Center for Light-Matter University, Tel Aviv
	University, Tel Aviv, Israel\\
	$^{3}$Instituto de Alta Investigaci\'{o}n, Universidad de Tarapac\'{a},
	Casilla 7D, Arica, Chile}

\begin{abstract}
We report results of systematic investigation of dynamics featured by moving
two-dimensional (2D) solitons generated by the fractional nonlinear Schr\"{o}%
dinger equation (FNLSE) with the cubic-quintic\ nonlinearity. The motion of
solitons is a nontrivial problem, as the fractional diffraction breaks the
Galilean invariance of the underlying equation. The addition of the
defocusing quintic term to the focusing cubic one is necessary to stabilize
the solitons against the collapse. The setting presented here can be
implemented in nonlinear optical waveguides emulating the fractional
diffraction. Systematic consideration identifies parameters of moving
fundamental and vortex solitons (with vorticities $0$ and $1$ or $2$,
respectively) and maximum velocities up to which stable solitons persist,
for characteristic values of the L\'{e}vy index which determines the
fractionality of the underlying model. Outcomes of collisions between 2D
solitons moving in opposite directions are identified too. These are merger
of the solitons, quasi-elastic or destructive collisions, and breakup of the
two colliding solitons into a quartet of secondary ones.
\end{abstract}

\maketitle

\section{Introduction}

In terms of abstract calculus, the concept of fractional derivatives was
introduced by Niels Henrik Abel in 1823 \cite{Abel}, and by Joseph Liouville
in 1832 \cite{Liouville}. In the modern mathematical literature \cite%
{Uchaikin}, widely accepted is the Caputo's definition of the derivative of
a non-integer positive order $\alpha $ \cite{Caputo},
\begin{equation}
\left( \frac{d}{dx}\right) ^{\alpha }\psi =\frac{1}{\Gamma \left( 1-\{\alpha
\}\right) }\int_{0}^{x}\frac{\psi ^{\left( n\right) }(\xi )dx}{\left( x-\xi
\right) ^{\left\{ \alpha \right\} }},  \label{cap}
\end{equation}%
where $\psi (x)$ is a generic real function, $\psi ^{(n)}$ is the usual
derivative of integer order $n\equiv \lbrack \alpha ]+1$, where $[\alpha ]$
and $\{\alpha \}\equiv \alpha -[\alpha ]$ are the integer and fractional
parts of $\left[ \alpha \right] $, and $\Gamma $ is the Gamma-function.

In physics, fractional derivatives had appeared in the context of \textit{\
fractional quantum mechanics} introduced by Laskin \cite{Lask1,Lask2}. The
subject of that theory is the quantization, by means of the Feynman's
path-integral technique \cite{Feynman}, of the dynamics of particles whose
classical stochastic motion is different from the usual Brownian form.
Instead, the motion is performed by random \textit{L\'{e}vy flights}, with
the mean distance $|\bar{x}|$ of the particle from the initial position ($%
x=0 $) growing with time (in one dimension, 1D) as%
\begin{equation}
|\bar{x}|\sim t^{1/\alpha }.  \label{flight}
\end{equation}%
Here, parameter $\alpha $, known as the \textit{L\'{e}vy index} (LI), takes
values%
\begin{equation}
0<\alpha \leq 2  \label{LI}
\end{equation}%
\cite{Mandelbrot}. The usual Brownian random walk, with $|\bar{x}|\sim \sqrt{
t}$, corresponds to the LI's limit value, $\alpha =2$, while Eq. (\ref%
{flight}) with $\alpha <2$ demonstrates that the $|\bar{x}|$ grows faster
than $\sqrt{t}$ at $t\rightarrow \infty $.

The Schr\"{o}dinger equation for the wave function $\Psi \left( x,t\right) $
of the particle moving by L\'{e}vy flights in 1D, in the presence of a
potential $V(x)$, was derived by Laskin \cite{Lask1,Lask2} in the following
scaled form:
\begin{equation}
i\frac{\partial \Psi }{\partial t}=\frac{1}{2}\left( -\frac{\partial ^{2}}{
\partial x^{2}}\right) ^{\alpha /2}\Psi +V(x)\Psi ,  \label{FSE}
\end{equation}%
see also Ref. \cite{GuoXu}. The fractional derivative, which represents the
kinetic-energy term in Eq. (\ref{FSE}), is different from the abstract
Caputo's definition (\ref{cap}). Instead, it naturally arises as the \textit{%
\ Riesz derivative} \cite{Riesz}, which is built as the superposition of the
direct and inverse Fourier transforms ($x\rightarrow p\rightarrow x$),
\begin{equation}
\left( -\frac{\partial ^{2}}{\partial x^{2}}\right) ^{\alpha /2}\Psi =\frac{%
1 }{2\pi }\int_{-\infty }^{+\infty }dp|p|^{\alpha }\int_{-\infty }^{+\infty
}d\xi e^{ip(x-\xi )}\Psi (\xi ).  \label{Riesz derivative}
\end{equation}%
In the framework of this definition, the action of the fractional
kinetic-energy operator, $\left( -\partial ^{2}/\partial x^{2}\right)
^{\alpha /2}$, in the Fourier representation amounts to the straightforward
multiplication by $|p|^{\alpha },$ with $|p|^{\alpha }$ being an obvious
Fourier-space counterpart of this operator.

In the 2D space, the fractional Schr\"{o}dinger equation for the wave
function $\Psi \left( x,y\right) $ is a straightforward extension of Eq. (%
\ref{FSE}) \cite{Lask2}:
\begin{equation}
i\frac{\partial \Psi }{\partial t}=\frac{1}{2}\left( -\frac{\partial ^{2}}{
\partial x^{2}}-\frac{\partial ^{2}}{\partial y^{2}}\right) ^{\alpha /2}\Psi
+V(x,y)\Psi .  \label{2D FSE}
\end{equation}%
In the Fourier space with wavenumbers $\left( p,q\right) $, the action of
the fractional operator, which represents the 2D kinetic energy, amounts to
the multiplication by $\left( p^{2}+q^{2}\right) ^{\alpha /2}$, the explicit
integral form of the operator in the coordinate space being%
\begin{equation}
\left( -\frac{\partial ^{2}}{\partial x^{2}}-\frac{\partial ^{2}}{\partial
y^{2}}\right) ^{\alpha /2}\Psi =\frac{1}{(2\pi )^{2}}\int \int dpdq\left(
p^{2}+q^{2}\right) ^{\alpha /2}\int \int d\xi d\eta e^{i\left[ p(x-\xi
)+iq(y-\eta )\right] }\Psi (\xi ,\eta ),  \label{2D operator}
\end{equation}%
cf. Eq. (\ref{Riesz derivative}).

Stationary solutions to Eq. (\ref{2D FSE}), with real energy eigenvalue $\mu
$, are looked for in the usual form,%
\begin{equation}
\Psi \left( x,y,t\right) =e^{-i\mu t}U\left( x,y\right) ,  \label{U}
\end{equation}%
where function $U$ satisfies the stationary equation,%
\begin{equation}
\mu U=\frac{1}{2}\left( -\frac{\partial ^{2}}{\partial x^{2}}-\frac{\partial
^{2}}{\partial y^{2}}\right) ^{\alpha /2}U+V(x,y)U,  \label{Ustat}
\end{equation}%
or the 1D reduction of Eq. (\ref{Ustat}). Function $U$ is real for the
ground-state solution, and complex for excited states carrying the vorticity
(angular momentum).

It is relevant to mention that, in the framework of the mean-field theory,
the dynamics of Bose-Einstein condensates in ultracold atomic gases is
governed by the Gross-Pitaevskii equation (GPE), i.e., the Schr\"{o}dinger
equation for the single-particle wave function augmented by the cubic term $%
\sim |\Psi |^{2}\Psi $ which represents the mean-field effect of
inter-particle collisions \cite{Pit-Str}. A challenging possibility is to
consider a condensate of bosonic particles obeying the fractional Schr\"{o}%
dinger equation (\ref{2D FSE}) or its one-dimensional reduction (\ref{FSE}).
While a consistent derivation of the respective fractional GPE remains an
open problem, it is expected that the equation may be obtained the following
scaled form:%
\begin{equation}
i\frac{\partial \Psi }{\partial t}=\frac{1}{2}\left( -\frac{\partial ^{2}}{
\partial x^{2}}-\frac{\partial ^{2}}{\partial y^{2}}\right) ^{\alpha /2}\Psi
+V(x,y)\Psi +\sigma |\Psi |^{2}\Psi ,  \label{FGPE}
\end{equation}%
with $\sigma =+1$ or $-1$ corresponding to the repulsive or attractive sign
of the inter-article interactions, respectively \cite{review}.

\section{The basic model}

Thus far, fractional quantum mechanics has not been observed experimentally.
An alternative possibility of the physical realization of the fractional Schr%
\"{o}dinger equation was proposed by Longhi \cite{Longhi}, following the
commonly known similarity of the Schr\"{o}dinger equation in quantum
mechanics and the paraxial propagation equation for the classical optical
field, with time $t$ replaced by the propagation distance, $z$ \cite{KA}.
The setup proposed for the realization of this possibility is based on the $%
4f$ (four-focal-lengths) structure implemented in an optical cavity \cite%
{Longhi}. The central element is a phase mask, placed in the middle
(Fourier) plane of the cavity. A lens decomposes the beam's transverse
structure into its Fourier components with wavenumbers $\left( p,q\right) $.
Then, the distance from the optical axis of the point at which each
component passes the phase mask is, roughly speaking, $R\sim \sqrt{
p^{2}+q^{2}}.$ The action of the effective fractional diffraction on the
Fourier transform $\hat{\Psi}\left( p,q\right) $ of the optical beam
amounts, according to Eq. (\ref{Riesz derivative}) or (\ref{2D operator}),
to the local phase shift
\begin{equation}
\hat{\Psi}\left( p,q\right) \rightarrow \hat{\Psi}\left( p,q\right) \exp %
\left[ i\left( p^{2}+q^{2}\right) ^{\alpha }Z\right] ,  \label{Z}
\end{equation}%
where $Z$ is the propagation distance over which the effect of the
fractional diffraction is accumulated. In the proposed setup, the
corresponding differential phase shift is provided by the design of the
phase mask. In addition to that, the effect corresponding to potential $%
V(x,y)$ in Eq. (\ref{2D FSE}) (or in its one-dimensional counterpart (\ref%
{FSE})) can be emulated by a specially shaped mirror added to the cavity
setup \cite{Longhi}.

The setting which is outlined above provides a single-stage transformation
of the optical beam, which is represented, essentially, by Eq. (\ref{Z}).
The continuous fractional Schr\"{o}dinger equation (\ref{2D FSE})\ or (\ref%
{FSE}) is then introduced as a result of averaging over many cycles of the
intra-cavity circulation of light.

Experimental implementation of the setup realizing the effectively
fractional diffraction has not yet been reported. However, a recent
experimental work has implemented the realization of a similarly devised
scheme which demonstrates the effective fractional group-velocity dispersion
in a fiber-laser cavity, in which the differential phase shift of decomposed
temporal spectral components, emulating the action of the fractional
dispersion, was provided by a computer-generated hologram, inserted as the
central phase mask \cite{Shilong}.

Once the effective fractional diffraction can be implemented by means of the
light propagation in the (1+1)- and (2+1)-dimensional spatial domains, with
the transverse dimensions 1 and 2, respectively (here, +1 stands for the
propagation direction $z$), it is natural to include nonlinearity of the
optical medium, and thus introduce the fractional nonlinear Schr\"{o}dinger
equation (FNLSE) \cite{proceedings,soliton2}. In the case of the cubic
(Kerr) nonlinearity \cite{KA}, its scaled form is tantamount to the
conjectured fractional GPE (\ref{FGPE}). Further, the presence of the
self-focusing nonlinearity suggests a possibility to predict the existence
of fractional solitons, as stable solutions of the FNLSEs \cite%
{soliton1,LDong,Chen,LDong2,review}. A crucially important problem for the
stability of such solitons is the possibility of the onset the \textit{wave
collapse} in the FNLSE. Indeed, an elementary estimate demonstrates that the
one-dimensional FNLSE with LI $\alpha $ and the cubic self-focusing term
gives rise to the \textit{critical collapse} (which is initiated by an input
whose norm exceeds a final critical value \cite{Berge,Fibich}) in the case
of $\alpha =1$, and to the \textit{supercritical collapse} (which may be
initiated by the input with an arbitrarily small norm) in the case of $%
\alpha <1$. Thus, the one-dimensional cubic FNLSE may give rise to stable
solitons in the interval of $1<\alpha \leq 2$, including the usual
(non-fractional) NLSE with $\alpha =2$.

The two-dimensional FNLSE with the cubic self-focusing term is problematic,
as it leads to the critical collapse in the usual case of $\alpha =2$ (with
the commonly known stationary solutions in the form of unstable \textit{\
Townes solitons} \cite{Townes,Berge,Fibich}), and to the supercritical
collapse, hence strongly unstable solitons, in the case of $\alpha <2$. A
solution of the instability problem for the 2D solitons is the use of the
cubic-quintic (CQ) nonlinearity. In that case, the trend towards the
collapse, driven by the cubic self-focusing, is arrested by the higher-order
quintic defocusing term. This option is a relevant one, as the combination
of cubic and quintic terms provides an accurate approximation for the
nonlinearity of various optical media, such as chalcogenide glasses \cite%
{chalco}, liquid carbon disulfide \cite{Boudebs}, colloidal suspensions of
metallic nanoparticles \cite{Albert}, and epsilon-near-zero waveguides \cite%
{ENZ}. The respective (2+1)-dimensional FNLSE for the paraxial evolution of
amplitude $u\left( x,y;z\right) $ of the electromagnetic wave is
\begin{equation}
i{\frac{\partial u}{\partial z}}={\frac{1}{2}}\left( -\frac{\partial ^{2}}{
\partial x^{2}}-\frac{\partial ^{2}}{\partial y^{2}}\right) ^{\alpha
/2}u-|u|^{2}u+|u|^{4}u,  \label{2+1}
\end{equation}%
where the fractional-diffraction operator is defined as per Eq. (\ref{2D
operator}), and coefficients in front of the self-focusing cubic and
defocusing quintic terms are set equal to $1$ by means of rescaling.

Stationary soliton solutions to Eq. (\ref{2+1}) with real propagation
constant $k>0$ are looked for as%
\begin{equation}
u(x,y;z)=e^{ikz}\phi \left( x,y\right) ,  \label{k}
\end{equation}%
where $\phi \left( x,y\right) $ is a localized solution of equation%
\begin{equation}
k\phi +\frac{1}{2}\left( -\frac{\partial ^{2}}{\partial x^{2}}-\frac{
\partial ^{2}}{\partial y^{2}}\right) ^{\alpha /2}\phi -|\phi |^{2}\phi
+|\phi |^{4}\phi =0.  \label{ODE}
\end{equation}%
Soliton solutions are characterized by the integral power,%
\begin{equation}
P=\int \int \left\vert u\left( x,y,\tau \right) \right\vert ^{2}dxdy.
\label{Power}
\end{equation}

In the case of the non-fractional diffraction, $\alpha =2$, Eq. (\ref{ODE})
gives rise to families of 2D solitons with embedded integer\ vorticity
(winding number), $m\geq 0$ ($m=0$ corresponds to the fundamental solitons).
In polar coordinates $\left( r,\theta \right) $, the vortex-soliton
solutions are looked for as $\phi \left( r,\theta \right) =a(r)\exp \left(
im\theta \right) $, where the real amplitude $a(r)$ satisfies the radial
equation $\left( a^{\prime \prime }+r^{-1}a^{\prime }-m^{2}r^{-2}a\right)
-2\left( ka-a^{3}+a^{5}\right) =0$, supplemented by the boundary conditions $%
a(r)\sim r^{m}$ at $r\rightarrow 0$, and $a(r)\sim r^{-1/2}\exp \left( -%
\sqrt{-2k}\right) $ at $r\rightarrow \infty $. For all winding numbers $%
m=0,\pm 1,\pm 2,...$, 2D vortex solitons exist in the interval of values of
the propagation constant
\begin{equation}
0<k<3/16,  \label{3/16}
\end{equation}%
in which the power takes values $0<P<\infty $. In the limit of $k\rightarrow
3/16$, the soliton develops an indefinitely broad flat-top shape, with the
inner amplitude approaching the limit value%
\begin{equation}
\left( \left\vert \phi \right\vert \right) _{\max }\left( k=3/16\right) =
\sqrt{3}/2.  \label{max}
\end{equation}%
The entire family of the fundamental solitons, with $m=0$, is completely
stable, while each family of the vortex solitons has its specific stability
interval, $k_{\min }^{(m)}<k<3/16$, where the power is large enough \cite%
{Michinel}. The stability interval shrinks with the increase of $m$, \textit{%
\ viz}., $k_{\min }^{(1)}\approx 0.1487,k_{\min }^{(2)}\approx
0.1618,k_{\min }^{(3)}\approx 0.1700$ \cite{Pego}.

Solutions of 1D FNLSEs with the CQ nonlinearity and, in some cases, with
potential terms, were studied in detail, including solitons \cite{ZengZeng
1D}-\cite{1D Stephanovich}, breathers \cite{Inc breathers}, Airy waves \cite%
{Airy 1D}, and modulational instability \cite{MI}. In 2D, systematic results
have been reported for solitons, including \textquotedblleft unconventional"
ones \cite{ZengZeng 2D OL,2D with PT}, solitary vortices \cite{Pengfei CQ
vortex} and necklace clusters \cite{Pengfei necklace}. As well as the
non-fractional CQ NLSE (with $\alpha =2$), its 2D fractional counterpart,
with $\alpha <2$, produces fundamental and vortex solitons in interval (\ref%
{3/16}); however, the difference is that, in the case of $\alpha <2$, the
solitons exist with the power exceeding a certain minimum value \cite%
{Pengfei CQ vortex} (for illustration, see Figs. \ref{fig1}(a), \ref{fig3}%
(a), and \ref{fig5}(a) below). Accordingly, curves $P(k)$ feature two
segments, one with $dP/dk>0$, which satisfies the celebrated
Vakhitov-Kolokolov (VK)\ criterion \cite{VK,Berge,Fibich}, and another one,
with $dP/dk<0$. The latter segment is definitely unstable, as it does not
comply with the VK criterion. The branch of the fundamental-soliton family ($%
m=0$) with $dP/dk>0$ is completely stable, while this condition is only
necessary but not sufficient for the stability of the vortex-soliton
families with $|m|\geq 1$, only their parts with the power large enough, $%
P>P_{\min }^{(m)}(\alpha )$, being stable \cite{Pengfei CQ vortex}. The
stability-threshold value, $P_{\min }^{(m)}(\alpha )$, grows with the
increase of the FNLSE's fractionality, i.e., with the decrease of LI $\alpha
$, thus making the creation of stable vortex solitons more difficult for
smaller $\alpha $.

The first subject of the present work, addressed below in Section 3, is the
creation of moving 2D solitons (actually, these are tilted spatial solitons)
as solutions of Eq. (\ref{2+1}) with winding numbers $m=0$, $1$, and $2$ and
coordinate $x$ replaced by $x$ by $x^{\prime }=x-cz$, where real $c$ is the
\textquotedblleft velocity" (actually, the spatial tilt). It is enough to
consider the tilt only along coordinate $x$, as the tilt in any other
direction is equivalent to it in the isotropic system. Accordingly, Eq. (\ref%
{2+1}) is rewritten in the moving (tilted) coordinates:%
\begin{equation}
i{\frac{\partial u}{\partial z}-ic}\frac{\partial u}{\partial x}={\frac{1}{2}
}\left( -\frac{\partial ^{2}}{\partial x^{2}}-\frac{\partial ^{2}}{\partial
y^{2}}\right) ^{\alpha /2}u-|u|^{2}u+|u|^{4}u.  \label{tilted}
\end{equation}%
Stationary solutions to Eq. (\ref{tilted}) are looked for in the usual form,
i.e., as per Eq. (\ref{k}), where function $\phi \left( x,y\right) $
satisfies the equation%
\begin{equation}
k\phi +ic\frac{\partial \phi }{\partial x}+\frac{1}{2}\left( -\frac{\partial
^{2}}{\partial x^{2}}-\frac{\partial ^{2}}{\partial y^{2}}\right) ^{\alpha
/2}\phi -|\phi |^{2}\phi +|\phi |^{4}\phi =0,  \label{phi}
\end{equation}%
Note that, unlike the quiescent (untitled) solitons with $c=0$, solutions of
Eq. (\ref{phi}) with $c>0$ cannot be real.

The essential difference of the FNLSE from its non-fractional counterpart,
with $\alpha =2$, is that the fractional diffraction operator (\ref{2D
operator}) (as well as its 1D counterpart (\ref{Riesz derivative})) breaks
the Galilean invariance, hence Eqs. (\ref{tilted}) and (\ref{phi}) cannot be
transformed back into the original ones (\ref{2+1}) and (\ref{ODE}),
respectively. A result reported below is that stable moving solitons exist
up to a certain maximum velocity, $c_{\max }$, which depends on LI $\alpha $
and propagation constant $k,$ as well as on winding number $m$. At $%
c>c_{\max }$, the fundamental 2D solitons do not exist, while the vortex
ones, with $m=1$ and $2$, may exist, but are unstable.

Once stable moving solitons are available, the next relevant problem, which
is addressed below in Section 4, is collisions between the solitons moving
(tilted) in opposite directions. It is demonstrated that outcomes of the
collisions between fundamental solitons may be quasi-elastic or strongly
inelastic, leading to merger of the solitons, or their decay, or, depending
on the parameters, breakup of the two colliding solitons into a set of four
ones. The latter outcome is also found, as a generic one, for collisions
between counterpropagating vortex solitons, with identical or opposite
winding numbers, alike, i.e., $m_{1}=m_{2}=1$, or $m_{1,2}=\pm 1$.

\section{Moving solitons and vortices}

\subsection{Fundamental solitons ($m=0$)}

Numerical solutions of Eq. (\ref{phi}) for stationary solitons have been
obtained by means of the modified squared-operator method \cite{Yang}. The
computational domain in the $\left( x,y\right) $ plane was taken as a square
of size $|x|,|y|<200$, discretized by sets of $2^{9}$ points along each
dimension. Then, stability of the so produced solutions was verified by
means of direct simulations of Eq. (\ref{tilted}).

\subsubsection{The moving fundamental solitons for the L\'{e}vy index (LI) $%
\protect\alpha =1.5$}

For the traveling (tilted) fundamental 2D solitons (with $m=0$), the results
are presented here for two characteristic LI\ values, $\alpha =1.0\ $and $%
1.5 $, which produce solutions of the FNLSE which are sufficiently different
from the well-known family of solutions obtained from Eq. (\ref{2+1}) with
the ordinary nonfractional diffraction, \ i.e., $\alpha =2$.

\begin{figure}[h]
\centering\includegraphics[width=5in]{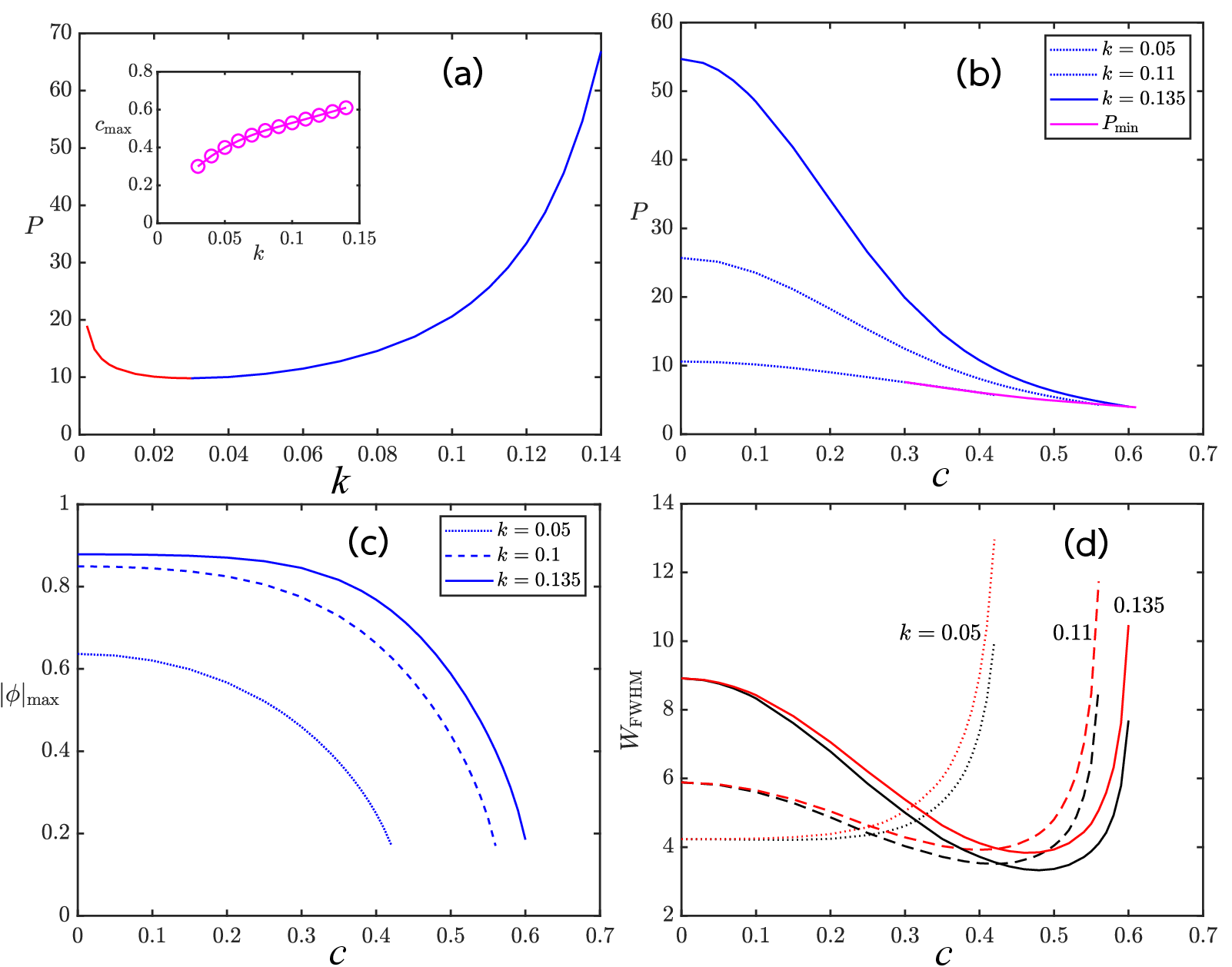}
\caption{Numerical results for the stable 2D fundamental solitons (with zero
vorticity) produced by Eq. (\protect\ref{phi}) with LI $\protect\alpha =1.5$
. (a) Power $P$ of the quiescent (untilted) solitons (corresponding to $c=0$
in Eq. (\protect\ref{phi})) vs. their propagation constant, $k$. The inset
shows the largest velocity (tilt) $c_{\max }$, up to which the moving
solitons exist for given $k$. Other panels show the power $P$ (b), amplitude
$|\protect\phi |_{\max }$ (c), and full width at half-maximum $W_{\mathrm{\
FWHM}}$ in the $x$ and $y$ directions (shown by the black and red curves,
respectively) (d), vs. the velocity (tilt) $c$, for different fixed values
of $k$. In panel (b), $P_{\min }$ $(c)$ is the smallest value of the power
for which the moving solitons are found for given $c$, considered as $%
c_{\max }$, with any value of $k$. }
\label{fig1}
\end{figure}

For $\alpha =1.5$, the summary of the numerically generated results is
presented in Fig. \ref{fig1}. The main plot in panel (a) reproduces, for the
reference's sake, the plot $P(k)$ for the family of the fundamental
quiescent (untitled, $c=0$) solitons, which is the same as reported (in a
different notation) in Ref. \cite{Pengfei necklace}. As mentioned above, the
subfamilies with $dP/dk>0$ and $dP/dk<0$ are, respectively, stable and
unstable, in full compliance with the VK criterion. The first set of new
results for the moving (tilted) 2D fundamental solitons is reported by means
of the inset in Fig. \ref{fig1}(a), \textit{viz}., the largest value $%
c_{\max }$ of velocity $c$ in Eq. (\ref{phi}) up to which the numerical
solution of this equation produces solitons. Detailed results for families
of the moving stable fundamental solitons are summarized in Figs. \ref{fig1}%
(b,c,d) by means of dependences of the power $P$, amplitude $\phi _{\max }$,
and widths $W_{\mathrm{FWHM}}$ in the direction of motion, $x$, and in the
perpendicular direction, $y$, on velocity, $c$, for several fixed values of
the propagation constant, $k$. In these figures, the curves terminate at
points $c=c_{\max }$ (in compliance with the inset in Fig. \ref{fig1}(a)),
beyond which numerical solutions for the moving solitons could not be found.
In addition, Fig. \ref{fig1}(b) shows the smallest value of the power, $%
P_{\min }$, up to which moving solitons are found for given $c$ (actually,
considered as $c_{\max }$) and any value of $k$.

\begin{figure}[h]
\centering\includegraphics[width=4in]{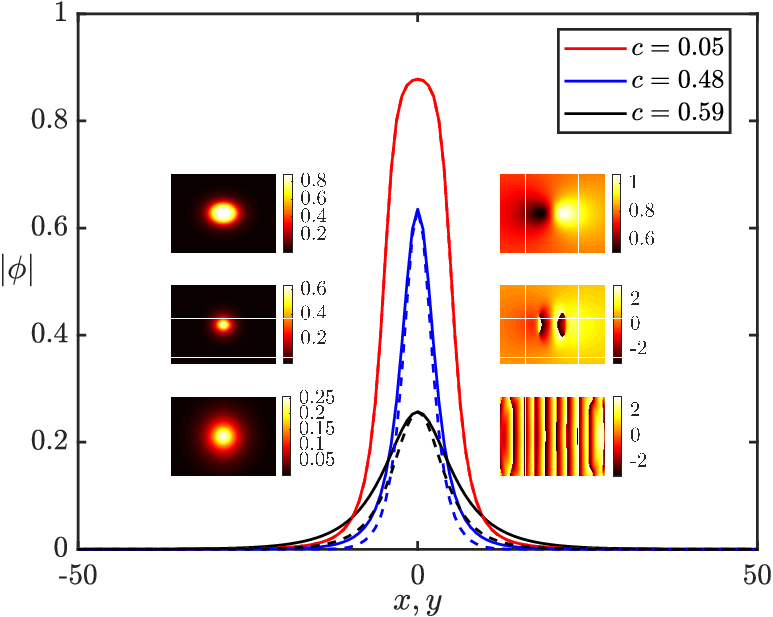}
\caption{Amplitude profiles of the stable moving (tilted) fundamental
solitons with fixed LI $\protect\alpha =1.5$ and propagation constant $%
k=0.135$, for different values of the velocity (tilt) $c$. The dashed and
solid lines represent the profiles along the $x$- and $y$-axes,
respectively. Insets display top views of the 2D amplitude and phase shapes
(the left and right plots, respectively). The top, middle, and bottom inset
rows correspond, severally, to $c=0.05,0.48$, and $0.59$. }
\label{fig2}
\end{figure}

Typical examples of stable moving solitons with $\alpha =1.5$ are displayed
in Fig. \ref{fig2}. Both Figs. \ref{fig1} and \ref{fig2} demonstrate that
the increase of the velocity (tilt) leads to monotonous decrease of their
amplitude and power. As concerns the width of the fundamental solitons, Fig. %
\ref{fig1}(d) demonstrates its nonmonotonous behavior, \textit{viz}.,
originally the solitons shrink with the increase of the velocity, changing
to expansion after reaching a minimum value of the width. In this
connection, it is relevant to note that the increase of the solitons' width
at $c\rightarrow c_{\max }$ makes it difficult to find them in a numerically
exact form. At small $c$, Fig. \ref{fig2} manifests the above-mentioned
trend to the formation of the fundamental solitons with the flat-top profile.

\begin{figure}[h]
\centering\includegraphics[width=2.3in]{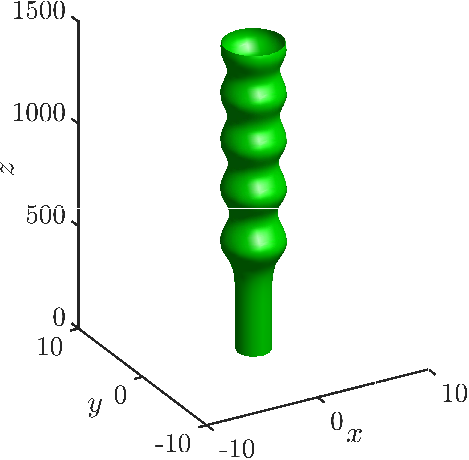}
\caption{The evolution of an unstable fundamental ($m=0$) soliton with $c=0$
, $\protect\alpha =1.5$, and $k=0.02$, produced by simulations of Eq. (
\protect\ref{2+1}) and displayed by means of the isosurface of the local
intensity, $|u(x,y)|^{2}=0.171$. The soliton spontaneously transforms into a
breather.}
\label{fig9}
\end{figure}

As concerns unstable solitons, which, in the case of $c=0$, populate the
segment of the $P(k)$ curve in Fig. \ref{fig1}(a) with $dP/dk<0,$ a typical
outcome of the development of the instability is spontaneous \
transformation of the stationary fundamental soliton into a robust breather,
as shown in Fig. \ref{fig9}. At $k<0.01$, the unstable fundamental soliton
does not evolve into a breather, but decays (not shown here in detail).

\subsubsection{The moving fundamental solitons for the L\'{e}vy index (LI) $%
\protect\alpha =1.0$}

\begin{figure}[h]
\centering\includegraphics[width=5in]{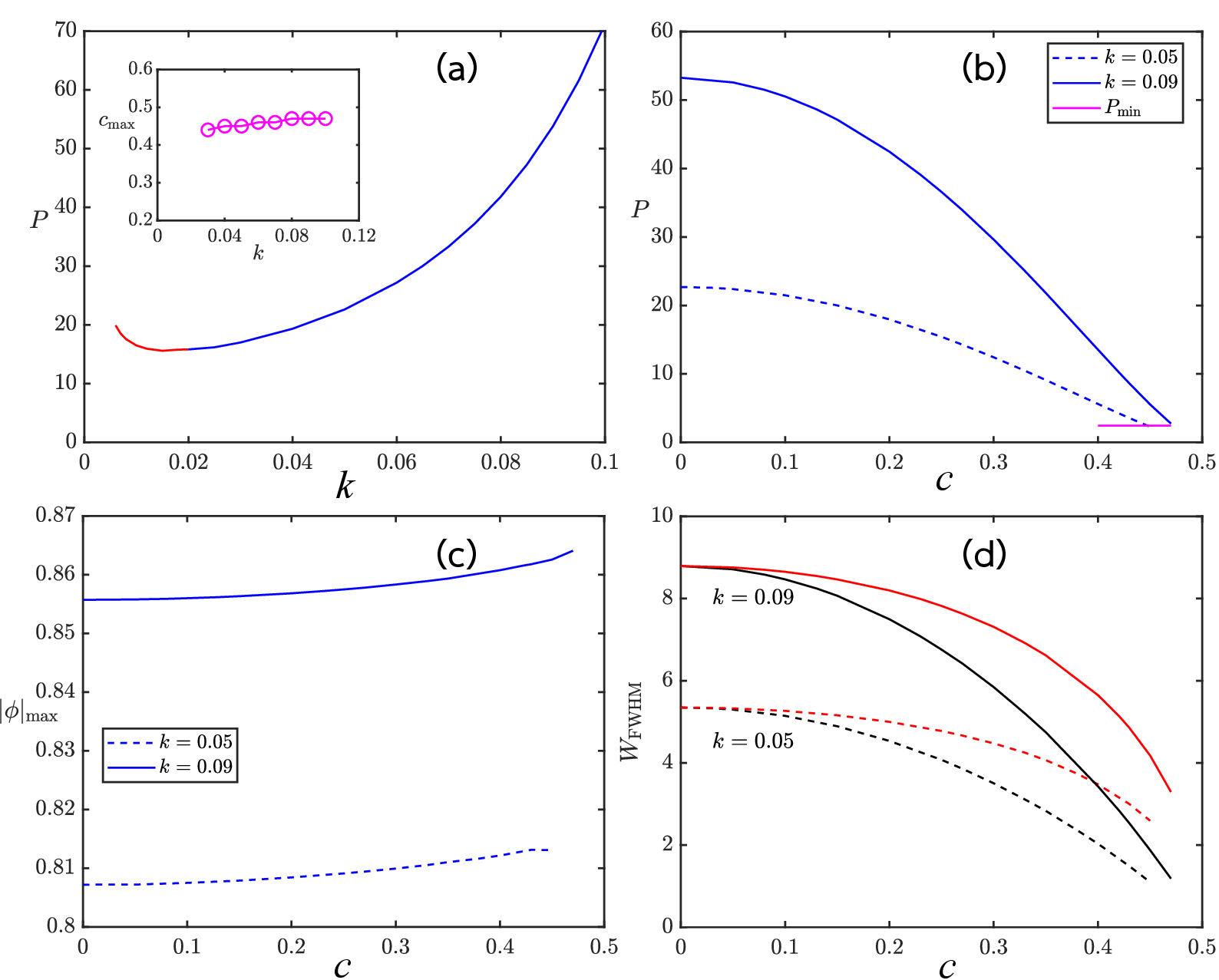}
\caption{The same as in Fig. \protect\ref{fig1}, but for the family of the
2D fundamental solitons found for LI $\protect\alpha =1.0$. The fixed values
of $k$ are indicated in panels (b) - (d).}
\label{fig3}
\end{figure}

For an essentially smaller value of the LI, $\alpha =1.0$, the results for
families of quiescent and moving 2D fundamental solitons are collected in
Fig. \ref{fig3}, following the pattern of Fig. \ref{fig1} for $\alpha =1.5$.
Here too, the curves plotted in panels (b)-(d) terminate at points $%
c=c_{\max }$ (in compliance with\ the inset to panel (a)), beyond which
solutions for the moving soliton could not be found. Further, it is seen
that the dependence of $c_{\max }$ on $k$ \ in the inset to Fig. \ref{fig3}%
(a), and the dependence of the power on the velocity (tilt) $c$, are similar
to those shown in Fig. \ref{fig1} for $\alpha =1.5$. However, the
dependences of the amplitude and width on $c$, observed in Figs. \ref{fig3}%
(c) and (d), are drastically different from the case of $\alpha =1$: the
amplitude does not decrease, but slowly increases with $c$, while the width
decreases monotonously, without the switch to increase (cf. Fig. \ref{fig1}%
(d) for $\alpha =1.5$).

\begin{figure}[h]
\centering\includegraphics[width=4in]{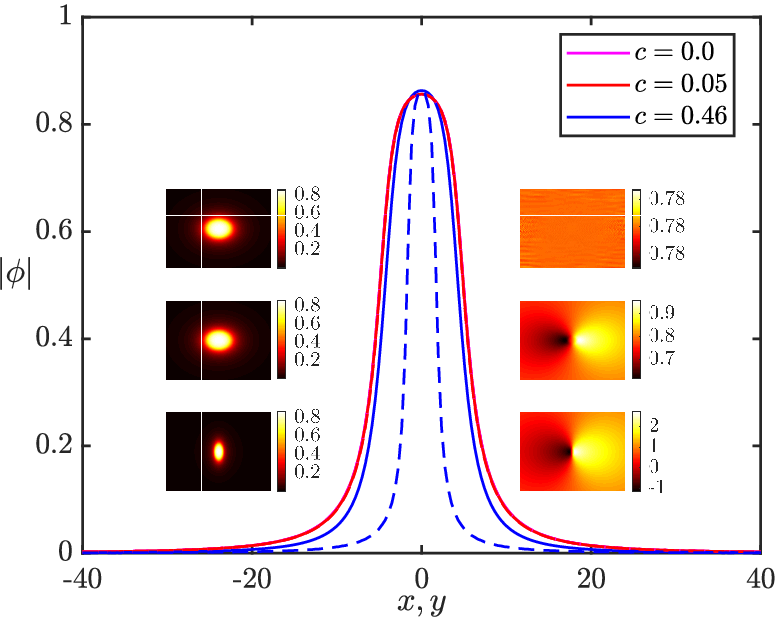}
\caption{The same as in Fig. \protect\ref{fig2}, but for LI $\protect\alpha %
=1.0$ and fixed propagation constant $k=0.09$.}
\label{fig4}
\end{figure}

Typical examples of stable moving (tilted) solitons, with different values
of velocity (tilt) $c$, for $\alpha =1.0$ and fixed propagation constant $%
k=0.09$, are displayed in Fig. \ref{fig4}. In comparison with the case of $%
\alpha =1.5$, the above-mentioned difference is that the soliton's amplitude
almost does not depend on $c$, while the shape keeps shrinking with the
increase of $c$. In this case, unstable solitons also tend to spontaneously
transform into oscillatory modes, cf. Fig. \ref{fig9} (not shown here in
detail).

\subsection{Vortices ($m=1$ and $\ 2$)}

For vortex solitons, systematic results have been collected for $m=1$ and $%
\alpha =1.5$. Stable vortex modes exist also for $m\geq 2$ at $\alpha =1.5$,
and for $m=1$ at $\alpha =1.0$, but they are stable at very large values of
the power \cite{Pengfei CQ vortex}. Because the inner amplitude of the modes
cannot exceed the maximum value (\ref{max}), this implies the necessity to
construct solutions in very large spatial domains, which makes the numerical
computations challenging.

\begin{figure}[h]
\centering\includegraphics[width=5in]{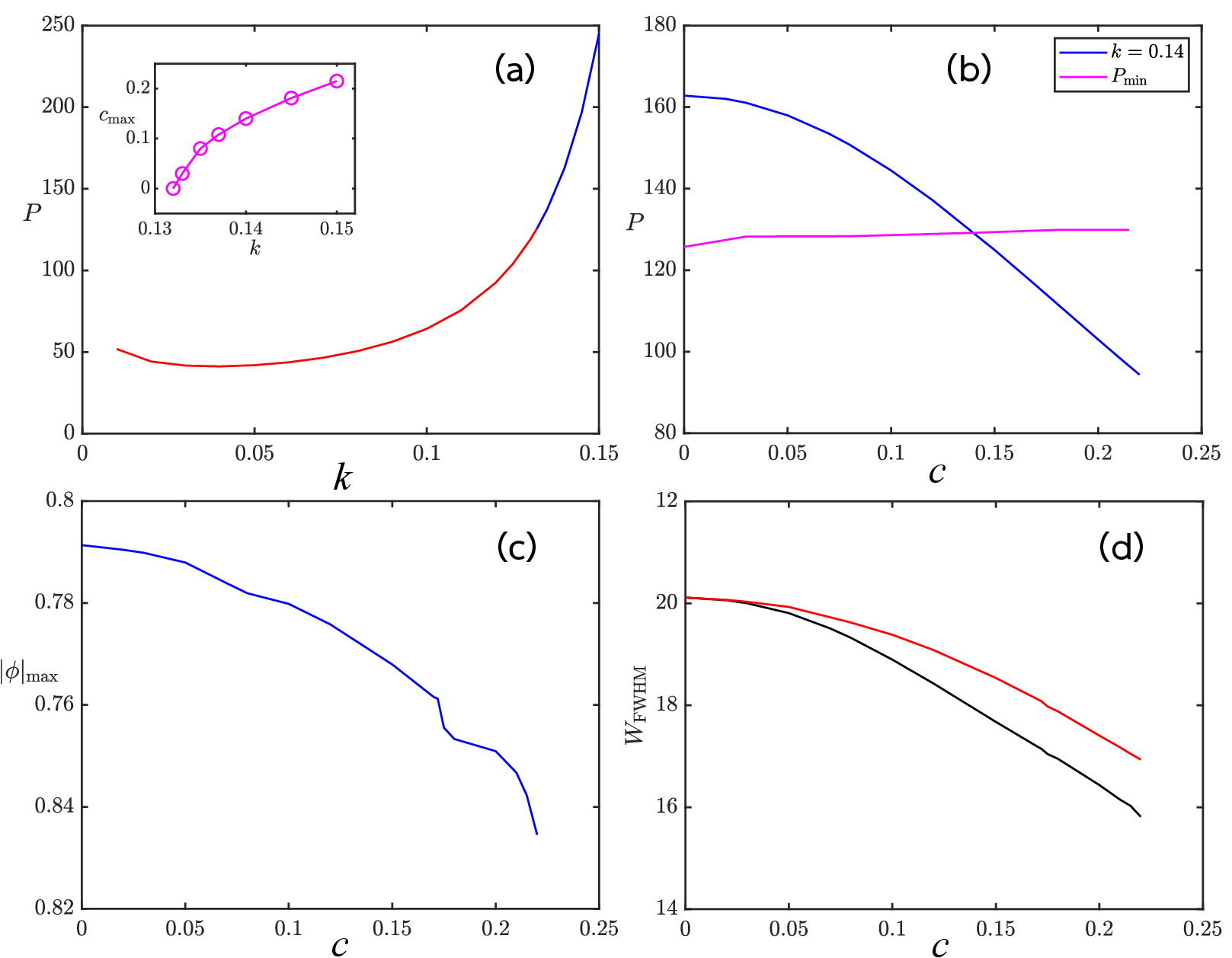}
\caption{(a) Power $P$ of the quiescent (untitled) vortex solitons
(corresponding to $c=0$ in Eq. (\protect\ref{phi})) with $m=1$ vs. the
propagation constant $k$. The inset shows the largest velocity (tilt) $%
c_{\max }$, up to which stable moving vortex solitons exist for given $k$.
Other panels show the power $P$ (b), amplitude $|\protect\phi |_{\max }$
(c), and full width at half-maximum $W_{\mathrm{FWHM}}$ in the $x$ and $y$
directions (shown by the black and red curves, respectively) (d), vs. the
velocity (tilt) $c$ of the vortex solitons, for a fixed propagation
constant, $k=0.14$. In panel (b), $P_{\min }$ $(c)$ is the smallest value of
the power for which the moving vortex solitons are stable for given $c$
(considered as $c_{\max }$), with any value of $k$.\ \ \ \ \ \ \ \ \ \ \ \ \
\ \ \ \ \ \ \ \ \ \ \ \ \ \ }
\label{fig5}
\end{figure}

Following the pattern of Figs. \ref{fig1} and \ref{fig3}, characteristics of
the family of the vortex solitons with $m=1$ and fixed propagation constant,
$k=0.14$, are presented in Fig. \ref{fig5}. Stable and unstable vortex
solitons populate the blue and red segments of the $P(k)$ curve in panel
(a). Unlike the families of the fundamental solitons (cf. Figs. \ref{fig1}%
(a) and \ref{fig3}(b)), Fig. \ref{fig5}(a) demonstrates that the VK
criterion is not sufficient for the stability of the vortex solitons,
because their instability against spontaneous splitting (see Fig. \ref{fig10}
below) is accounted for by complex eigenvalues, which are not taken into
regard by the VK condition \cite{Pego,Pengfei CQ vortex}.

Overall, the dependences $P(c)$ and $|\phi |_{\max }(c)$ for the vortex
solitons, plotted in Figs. \ref{fig5}(b) and (c), are qualitatively similar
to the same dependences for the fundamental solitons in Figs. \ref{fig1}(b)
and (c), while the dependence of the widths on $c$ in Fig. \ref{fig5}(d) is
a monotonously decreasing one, in contrast with its nonmonotonous
counterpart for the fundamental solitons at the same value of the LI, $%
\alpha =1.5$. Also different are the dependences $c_{\max }(k)$ for the
fundamental and vortex solitons: as one sees from the comparison of insets
in Figs. \ref{fig1}(a) and \ref{fig5}(a), this dependence is strong in the
former case, and very weak in the latter one.

The meaning of the value $P_{\min }(c)$ for the moving vortex solitons,
which is shown in panel (b), is different from that of $P_{\min }(c)$ for
the fundamental solitons (cf. Figs. \ref{fig1}(b) and \ref{fig3}(b)): while
the fundamental solitons do not exist at $P<P_{\min }$, their vortex
counterparts may exist at $P<P_{\min }$, but they are unstable against
spontaneous splitting, as shown below in Fig. \ref{fig11}.

A peculiarity of the dependence of the amplitude on $c$ for the vortex
soliton, observed in Fig. \ref{fig5}(c), is that it is non-smooth, showing,
in particular, a conspicuous irregularity around $c=0.172$. However, this
feature is not a significant one, as it belongs to the unstable segment of
the family, according to Fig. \ref{fig5}(b).{\LARGE \ }

\begin{figure}[h]
\centering\includegraphics[width=4in]{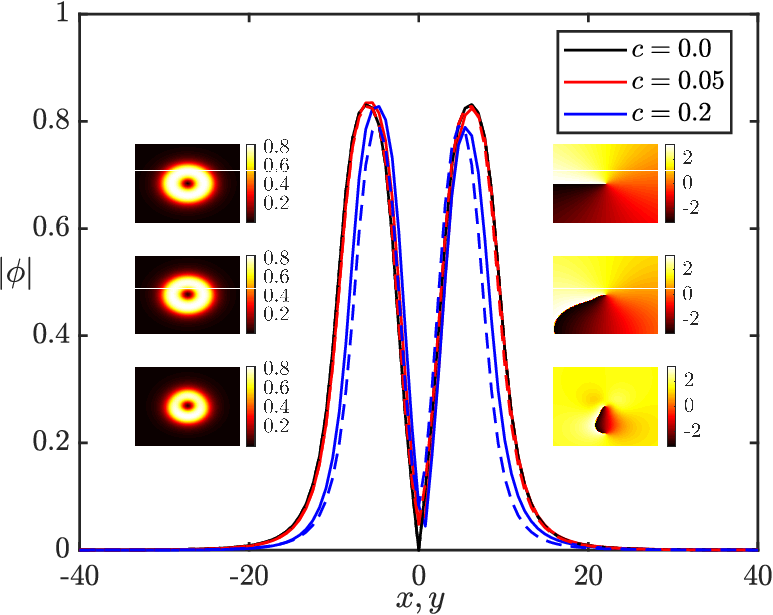}
\caption{Amplitude profiles of moving vortex solitons for the fixed
propagation constant, $k=0.14$, and different velocities. Dashed and solid
lines represent the profiles in the $x$ and $y$ directions, respectively.
The left and right inset panels are 2D plots of the amplitude and phase
shapes. Top and middle rows represent stable vortex solitons with $c=0$ and $%
c=0.05$, respectively. The bottom row, with $c=0.2$, corresponds to an
unstable vortex soliton, in agreement with Fig. \protect\ref{fig5}(b) (the
power of this solutions is $P=103.01<P_{\min }\approx 129.17$). The unstable
soliton spontaneously breaks up in two separating fragments, which are close
to fundamental solitons, see Fig. \protect\ref{fig11}.}
\label{fig6}
\end{figure}

Typical examples of stable and unstable quiescent ($c=0$) and moving ($c>0$)
vortex solitons are displayed in Fig. \ref{fig6}. In addition, an example of
the splitting instability of a quiescent\ vortex soliton with $k=0.1$, which
indeed belongs to the unstable subfamily in Fig. \ref{fig5}(a)), is
displayed in Fig. \ref{fig10}. It breaks up into a set of two separating
fragments, each one being close to a fundamental soliton. The picture
demonstrates conversion of the spin momentum of the original vortex into the
orbital angular momentum of the moving fragments. In addition to that, each
secondary soliton exhibits weak elliptical deformation and inner spinning,
which also carries a small fraction of the total momentum.

Moving vortices with $P<P_{\min }$ are unstable too, breaking up in two
quasi-soliton fragments, which separate in the $y$ direction (perpendicular
to the velocity), as shown for $k=0.14$ and $c=0.2$ in Fig \ref{fig11}. The
stationary profile of this unstable vortex soliton is displayed in Fig.\ \ref%
{fig6}.

\begin{figure}[h]
\centering\includegraphics[width=2.3in]{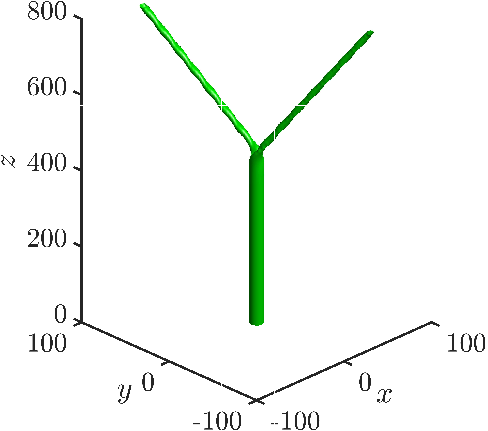}
\caption{The evolution of an unstable vortex soliton with $\protect\alpha %
=1.5$, $c=0$ and $k=0.1$, as produced by simulations of Eq. (\protect\ref%
{2+1}), is shown by means of the isosurface of the local intensity, $%
|u(x,y)|^{2}=0.357$. The vortex spontaneously splits in two obliquely moving
modes which are close to stable fundamental solitons.}
\label{fig10}
\end{figure}

\begin{figure}[h]
\centering\includegraphics[width=2.3in]{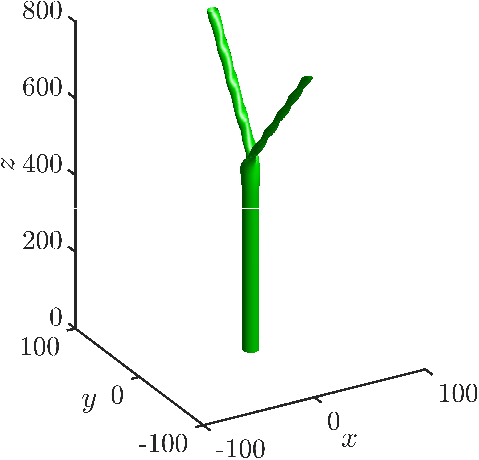}
\caption{The evolution of an unstable vortex soliton with $\protect\alpha %
=1.5$, $c=0.2$ and $k=0.14$, as produced by simulations of Eq. (\protect\ref%
{tilted}) and shown by means of the isosurface of the local intensity $%
|u(x,y)|^{2}=0.414$. The power of this soliton is $P=103.01<P_{\min }\approx
129.17$, see Fig. \protect\ref{fig5}(b). The instability splits the vortex
in the set of two localized modes moving in the $y$ direction (perpendicular
to the velocity $c$), which are close to stable fundamental solitons.}
\label{fig11}
\end{figure}

\begin{figure}[h]
\centering\includegraphics[width=2.3in]{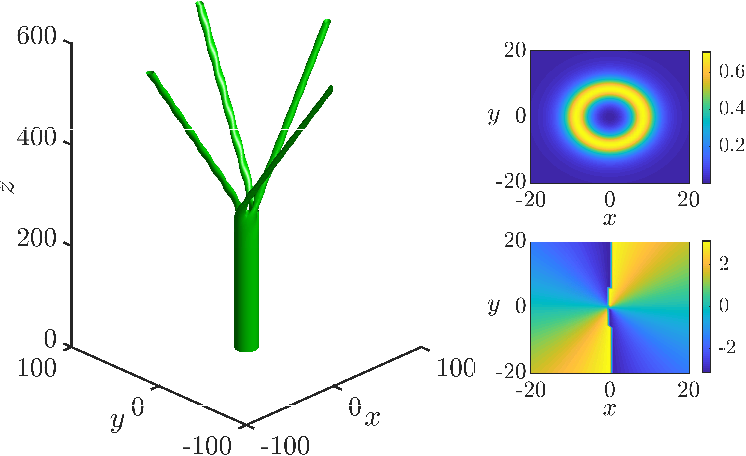}
\caption{The same as in Fig. \protect\ref{fig10}, but for an unstable
quiescent double-vortex soliton ,with $m=2$, $\protect\alpha =1.5$, $c=0$
and $k=0.1$, as produced by simulations of Eq. (\protect\ref{2+1}). The
spontaneous splitting of the vortex into the set of separating
quasi-solitons is shown by means of the isosurface of the local intensity, $%
|u(x,y)|^{2}=0.355$. The right top and bottom plots display the initial
amplitude and phase profiles of the vortex soliton.}
\label{fig16}
\end{figure}

For the sake of the completeness of the presentation, an example of the
instability of the quiescent double-vortex soliton (with $m=2$) is displayed
in Fig. \ref{fig16}. As is typical for the instability of vortex solitons
with the double winding number \cite{Pego}, the outcome is its fission into
the set of four separating fragments, each one being a spinning elliptically
deformed fundamental soliton (cf. the spontaneous fission of the unstable
vortex with $m=1$ into the set of two quasi-soliton fragments, exhibited in
Fig. \ref{fig6}).

\begin{figure}[h]
\centering\includegraphics[width=2.3in]{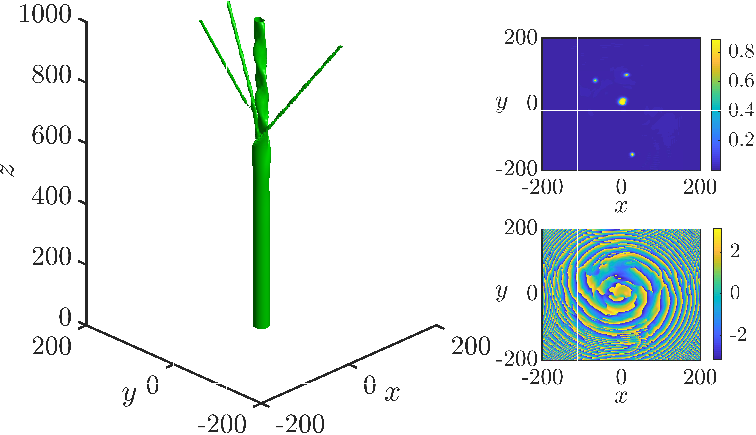}
\caption{An example of the splitting instability of a moving vortex soliton
with $m=2$, $c=0.03$, $k=0.14$, as produced by simulations of Eq. (\protect
\ref{2+1}) with $\protect\alpha =1.5$. The spontaneous separation of
low-power jets from the persistent central quasi-solitons is shown by means
of the isosurface of the local intensity, $|u(x,y)|^{2}=0.415$. The right
top and bottom plots display the final amplitude and phase profiles of the
optical field, at $z=1000$.}
\label{fig17}
\end{figure}

In all the simulations performed in this work, moving double-vortex
solitons, with $m=2$, demonstrate splitting instability. However, it is
different from the spontaneous fission of the quiescent vortex with $m=2$
into the set of four quasi-solitons with equal powers (cf. Fig. \ref{fig11}%
): as shown in Fig. \ref{fig17}, three low-power jets split off from the
central beam, which keeps a major share of the initial power, in the form of
a deformed quasi-solitons featuring inner spinning.

\section{Collisions between moving solitons}

The availability of stably moving fundamental and vortex solitons suggests a
possibility to study collisions between identical ones moving in opposite
directions, $\pm x$. To this end, Eq. (\ref{2+1}) was simulated, with the
input in the form of a pair of identical solitons produced by numerical
solutions of Eq. (\ref{phi}) with equal vorticities $m=0$ or $1$ and
opposite velocities $\pm c$. Centers of the two solitons are initially
placed at points $x=\pm x_{0}$, with a sufficiently large separation $2x_{0}$%
.

\subsection{Collisions between fundamental solitons}

\begin{figure}[h]
\centering\includegraphics[width=4in]{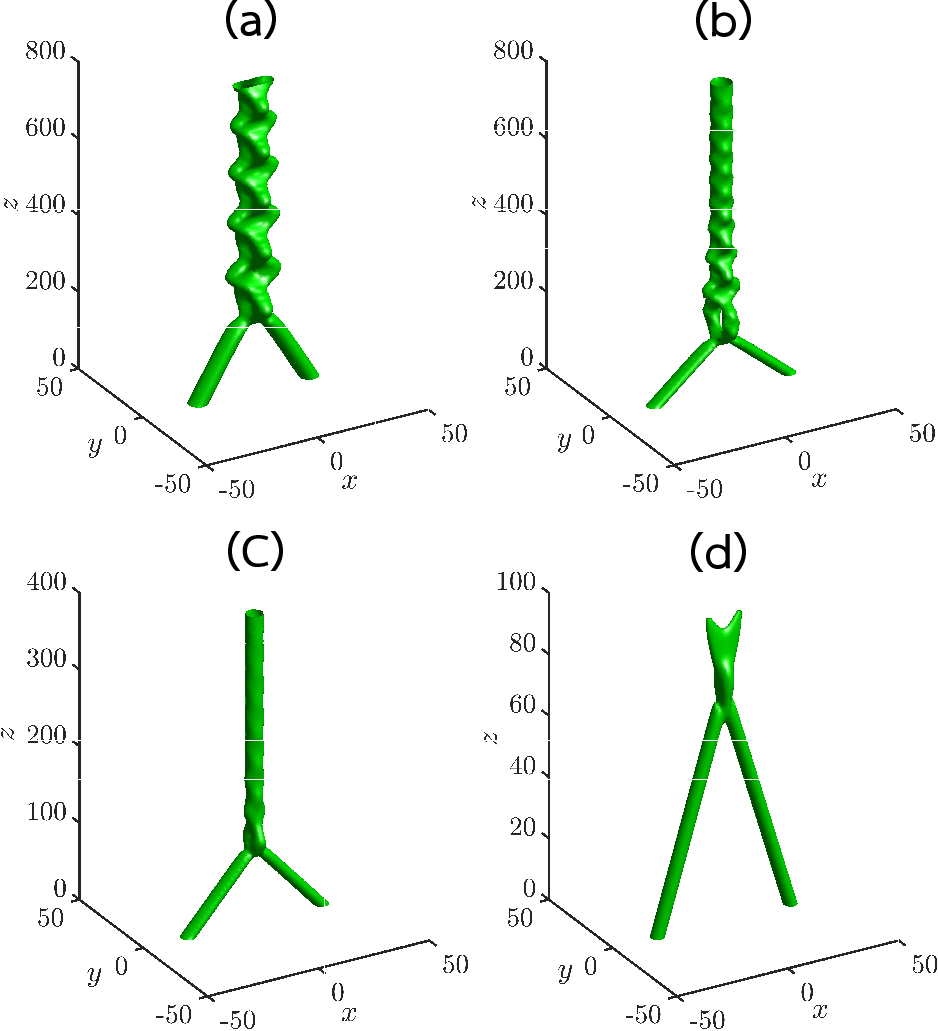}
\caption{Outcomes of collisions between moving fundamental solitons with the
propagation constant $k=0.135$ and velocities $c=\pm 0.1$ (a), $\pm 0.2$
(b), $\pm 0.3$ (c), and $\pm 0.4$ (d). The results are produced by
simulations of Eq. (\protect\ref{2+1}) with $\protect\alpha =1.5$, and are
shown by means of the isosurface of the local intensity (a) $%
|u(x,y)|^{2}=0.336$; (b) $|u(x,y)|^{2}=0.392$; (c) $|u(x,y)|^{2}=0.370$; (d)
$|u(x,y)|^{2}=0.113$; . The outcomes of the collision are merger of the
fundamental solitons into robust quiescent breathers in (a) and (b), merger
into a stationary quiescent soliton in (c), and destruction of the colliding
solitons in (d). The latter outcome is observed in interval (\protect\ref%
{decay}) of the collision velocities.}
\label{fig7}
\end{figure}

First, setting $\alpha =1.5$ in Eq. (\ref{2+1}), we present results of the
interaction between moving fundamental solitons with a fixed propagation
constant, $k=1.35$, and gradually increasing collision velocity $|c|$. For
small values of the velocity, Figs. \ref{fig1}(b,d) and \ref{fig2}
demonstrate that the moving solitons have higher powers and relatively large
widths. In this case, the simulations displayed in Fig. \ref{fig7} reveal
that the strong nonlinearity, caused by the large powers, leads to strongly
inelastic outcomes of the collisions: the solitons colliding with velocities
$c=\pm 0.1$ and $\pm 0.2$ (Figs. \ref{fig7}(a,b)) merge into quiescent
(standing) breathers, with the amplitude of inner oscillations decreasing
with the increase of $|c|$. As an extension of this trend, Fig. \ref{fig7}%
(c) demonstrates that the collision with velocities $c=\pm 0.3$ leads to the
merger of the colliding solitons into a single quiescent quasi-soliton,
without conspicuous oscillations.

Next, in the interval of the collision velocities
\begin{equation}
0.38\leq |c|\leq 0.45  \label{decay}
\end{equation}%
the weaker nonlinearity, corresponding to the smaller power and amplitude
(see Figs. \ref{fig1}(b) and (c)), is not sufficient to maintain the soliton
into which the colliding ones attempt to merge, therefore it quickly decays,
as shown in Fig. \ref{fig7}(d) for the collision with $c=\pm 0.40$. Thus,
the collisions are completely destructive in interval (\ref{decay}). Note
that the right edge of this interval is close to $c\approx 0.48$, at which
the soliton's width attains a minimum, $\left( W_{\mathrm{FWHM}}\right)
_{\min }\left( \alpha =1.5,k=1.35\right) \approx 3.325$, in Fig. \ref{fig1}%
(d).

\begin{figure}[h]
\centering\includegraphics[width=4in]{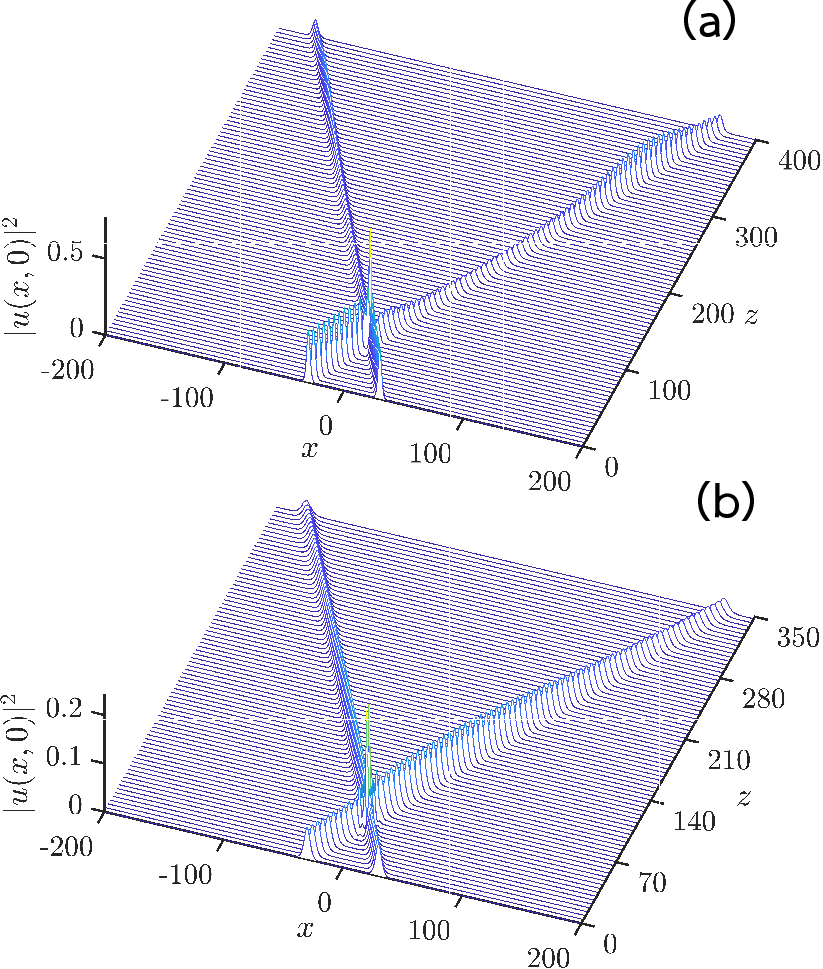}
\caption{Quasi-elastic collisions between the moving fundamental solitons
with $\protect\alpha =1.5$ and $k=0.135$, shown by means of the evolution of
the local power (intensity) in cross section $y=0$ for collision velocities $%
c=\pm 0.50$ (a) and $c=\pm 0.59$ (b).}
\label{fig8}
\end{figure}

For the values of LI $\alpha =1.50$ and propagation constant $k=1.35$ which
are considered here, the inset in Fig. \ref{fig1}(a) demonstrates that the
moving fundamental solitons exist up to $c_{\max }\left( \alpha
=1.5,k=1.35\right) \approx 0.6$. In the interval of the collision velocities
between the right edge of region (\ref{decay}) and $c_{\max }$, the
relatively large velocity and weak nonlinearity make the collisions
quasi-elastic, as shown in Figs. \ref{fig8}(a) and (b) for $|c|=0.50$ and $%
0.59$, respectively, the latter value being very close to $c_{\max }$.

In simulations of collisions between fundamental solitons with smaller
propagation constants (e.g., $k=0.05$), hence much lower powers and
amplitudes (see Figs. \ref{fig1}(b) and (c)), the merger does not occur. In
most cases, the collisions are destructive, becoming quasi-elastic at $|c|$
close to the respective value $c_{\max }$ ( not shown here in detail).

\begin{figure}[h]
\centering\includegraphics[width=2.3in]{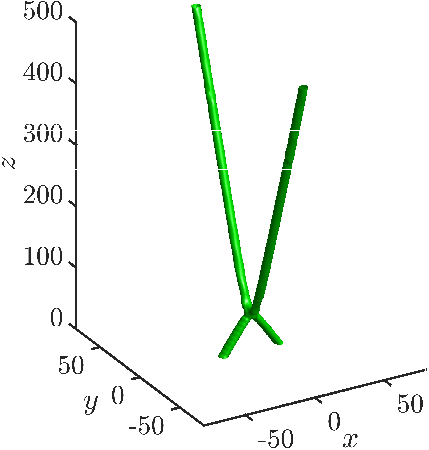}
\caption{The quasi-elastic collision between fundamental solitons with $%
c=\pm 0.3$ for $\protect\alpha =1.0$ and $k=0.09$, shown by means of the
isosurface of the local intensity, $|u(x,y)|^{2}=0.3419$. After the
collision, the solitons move in the perpendicular direction. The collisions
follow this scenario in interval (\protect\ref{c}).}
\label{fig12}
\end{figure}

In the case of LI $\alpha =1.0$, we take $k=0.09$ as the value of the
propagation constant corresponding to high power of the fundamental soliton,
as suggested by Fig. \ref{fig3}(a). In this case, the simulations
demonstrate merger of the colliding fundamental solitons into a breather at
relatively small velocities. In the velocity interval
\begin{equation}
0.21\leq |c|\leq 0.36,  \label{c}
\end{equation}%
the simulations reveal quasi-elastic collisions in a specific form: as shown
in Fig. \ref{fig12} for velocities $c=\pm 0.3$, the soliton colliding in the
$x$-direction separate, after the collision, along the $y$ axis, i.e., the
velocity vectors rotate by $90^{\mathrm{o}}$ in the $\left( x,y\right) $
plane.

\begin{figure}[h]
\centering\includegraphics[width=4in]{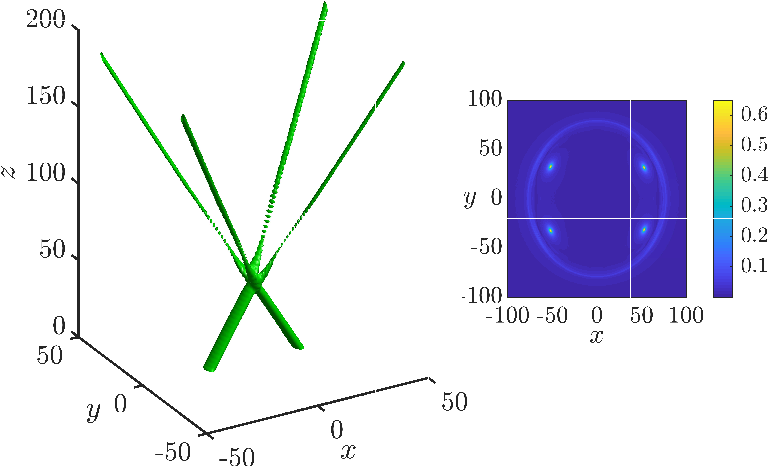}
\caption{(Left) The inelastic collision between the fundamental solitons
with $\protect\alpha =1.0$, $k=0.09$, and velocities $c=\pm 0.4$, shown by
means of the isosurface of the local intensity, with $|u(x,y)|^{2}=0.216$.
(Right) The intensity profile at $z=200$, which shows the set of four
secondary solitons produced by the collision. This outcome of the collisions
occurs in the interval of velocities $0.36<|c|<c_{\max }\approx 0.49$.}
\label{fig13}
\end{figure}

At $\alpha =1.0$, $k=0.09$, and $|c|>0.36$, up to $|c|=c_{\max }\approx 0.49$
(see Fig. \ref{fig3}(b)), collisions between fundamental solitons are
inelastic. The result, displayed in Fig. \ref{fig13} for $c=0.4$,
demonstrates breakup of the colliding pair into a set of four secondary
solitons, which move in oblique directions in the $\left( x,y\right) $ plane.

\subsection{Collisions between vortices ($m_{1}=m_{2}=1$ and $m_{1}=-m_{2}=1$
)}

\begin{figure}[h]
\centering\includegraphics[width=4in]{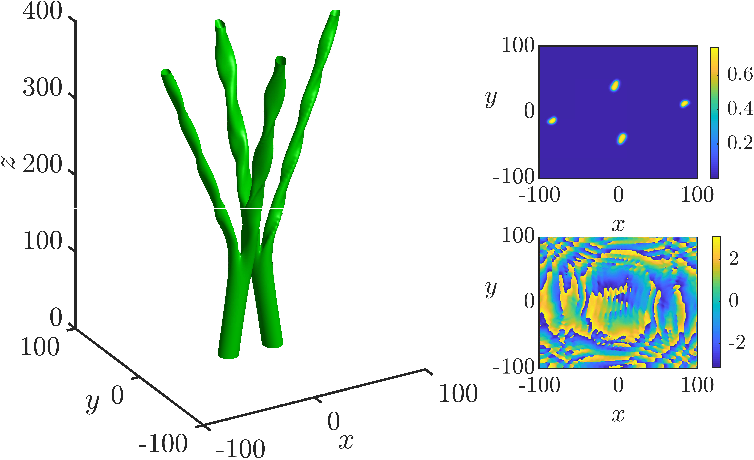}
\caption{(Left) The collision between two identical vortex solitons with
winding numbers $m=1$, propagation constant $k=0.14$ and velocities $c=\pm
0.1$, produced by simulations of Eq. (\protect\ref{2+1}) with $\protect%
\alpha =1.5$. The result is shown by means of the isosurface of the local
intensity $|u(x,y)|^{2}=0.380$. (Right) The intensity and phase profiles at $%
z=400$, which show the set of four secondary quasi-solitons produced by the
collision. The emerging elliptically deformed quasi-solitons rotate
counter-clockwise, keeping the original angular momentum of the colliding
vortices. This example illustrates the outcome of the collision between
counterpropagating identical vortex solitons observed in the generic case. }
\label{fig14}
\end{figure}

Similar to what is shown in Fig. \ref{fig13}, collisions between identical
counterpropagating vortex solitons with $\alpha =1.5$ and winding numbers $%
m=1$ lead, in \ the generic case, to their breakup into a set of four
secondary quasi-solitons. An example for the colliding solitary vortices
with propagation constant $k=0.14$ (these vortices are stable, according to
Fig. \ref{fig5}(a)), is displayed in Fig. \ref{fig14} for velocities $c=\pm
0.1$. The emerging quartet of quasi-solitons feature non-circular shapes,
each spinning counter-clockwise in the course of the propagation, to keep
the angular momentum of the original vortices.

It is natural too to explore collisions between counter-rotating vortex
solitons, with winding numbers $m=\pm 1$. The result is shown in Fig. \ref%
{fig15}, for the same values of the parameters, $k=0.14$, $c=\pm 0.1$, and $%
\alpha =1.5$, as in Fig. \ref{fig14}. The outcome is again fission of the
colliding vortices into a set of four separating quasi-solitons.
Nevertheless, an essential difference is that the lack of the necessity to
keep the initial total angular moment, which is zero in this case, allows
all the emerging quasi-solitons to move in the direction of the original
collision (along the $x$ axis), while in Fig. \ref{fig14} two of them are
moving, essentially, in the perpendicular direction.

\begin{figure}[h]
\centering\includegraphics[width=4in]{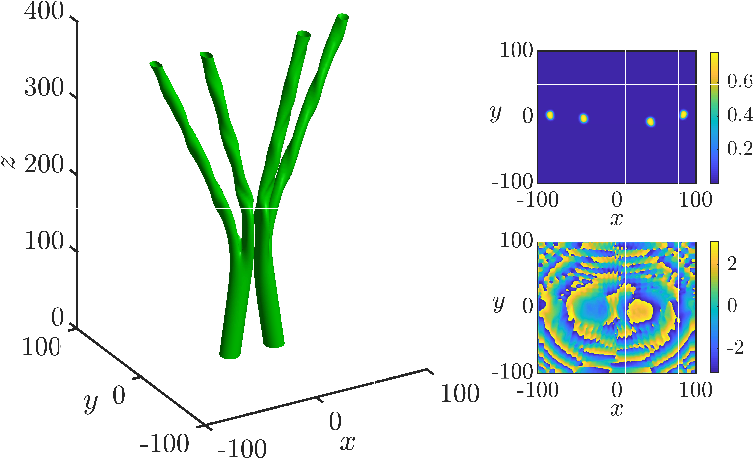}
\caption{The same as in Fig. \protect\ref{fig14}, but for the collision of
the vortex solitons with winding numbers $m=+1$ and $-1$.}
\label{fig15}
\end{figure}

\section{Conclusion}

The objective of this work is to develop systematic analysis of dynamics
initiated by motion of localized modes produced by the two-dimensional FNLSE
(fractional nonlinear Schr\"{o}dinger equation) with the CQ (cubic-quintic)
nonlinearity. The motion problem is nontrivial for the FNLSE because the
fractional diffraction does not admit Galilean invariance of the equation.
The choice of the nonlinearity with competing self-focusing cubic and
defocusing quintic terms is necessary to stabilize the localized modes, in
the form of fundamental\ (zero-vorticity) and vortex solitons, against the
collapse (however, the stability of the vortex solitons against spontaneous
splitting into sets of fundamental quasi-solitons remains a nontrivial
issue). The model can be implemented in optical cavities which combine the
emulation of the fractional diffraction and CQ material nonlinearity.

By means of the systematic numerical analysis, parameters of stably moving
fundamental and vortex solitons (with winding numbers $m=0$ and $1$,
respectively) and largest velocities, up to which the\ stable solitons
exist, are identified, for characteristic values of the LI (L\'{e}vy index)
which defines the fractionality of the underlying FNLSE. Then, collisions
between counterpropagating 2D solitons are explored by means of numerical
simulations. The results demonstrate different outcomes of the collisions,
depending on parameters, \textit{viz}., merger of colliding solitons,
quasi-elastic and destructive collisions, and the breakup of the colliding
pairs into sets of four quasi-solitons. In the latter case, the colliding
vortex solitons split into a quartet of spinning quasi-solitons.

The analysis reported in this work can be naturally extended in other
directions. One of them is the consideration of the fractional 2D complex
Ginzburg-Landau equation with the cubic-quintic nonlinearity, which may
produce a variety of stable localized modes \cite%
{CQ-CGL-1D,Yingji-CGL,Yingji-elliptic-CGL,discrete-CGL}.

\section*{Acknowledgements}

The Work of T.M. is supported by Faculty of Engineering, Naresuan University
through No. The work of B.A.M. is supported, in paper, by Israel Science
Foundation through grant No. 1695/22.

\end{document}